# Cascading Citation Expansion

Chaomei Chen[a,b]
[a]College of Computing and Informatics, Drexel University, Philadelphia PA 19104-2875, USA
[b]Department of Information Science, Yonsei University, Seoul, South Korea
Email: chaomei.chen@drexel.edu

## Abstract
Digital Science's Dimensions is envisaged as a next-generation research and discovery platform for a better and more efficient access to cross-referenced scholarly publications, grants, patents, and clinical trials. As a new addition to the growing open citation resources, it offers opportunities that may benefit a wide variety of stakeholders of scientific publications from researchers, policy makers, and the general public. In this article, we explore and demonstrate some of the practical potentials in terms of cascading citation expansions. Given a set of publications, the cascading citation expansion process can be iteratively applied to a set of articles so as to extend the coverage to more and more relevant articles through citation links. Although the conceptual origin can be traced back to Garfield's citation indexing, it has been largely limited, until recently, to the few who have unrestricted access to a citation database that is large enough to sustain such iterative expansions. Building on the open API of Dimensions, we integrate cascading citation expansion functions in CiteSpace and demonstrate how one may benefit from these new capabilities. In conclusion, cascading citation expansion has the potential to improve our understanding of the structure and dynamics of scientific knowledge.
## Keywords
Scientometrics, open citations, Dimensions, CiteSpace, cascading citation expansion

## Introduction

One of the grand challenges for researchers and stakeholders of science and innovation is the accessibility to scientific knowledge at various stages not only in terms of searching and reading scholarly publications but also understanding concrete claims and their significance (C. Chen, 2016). In addition to the accessibility challenges, the integral role of meta-knowledge in science has been increasingly realized. Meta-knowledge, the knowledge of scientific knowledge, is what we know about the epistemic status of scientific claims, hypotheses, consensus, controversies, and various uncertainties (C. Chen & Song, 2017). The idea of citation indexing was originally proposed by Eugene Garfield as a remedy to the vocabulary mismatch problem in the traditional keyword-based search (E. Garfield, 1963; Eugene Garfield, Sher, & Torpie, 1964). Two articles are likely to be relevant if one cites the other despite they may have no words in common. Thus if we know one article is relevant, then it is possible to find additional articles that are likely to be relevant as well. Finding relevant articles based on the existence of citation links is known as citation expansion. Citation expansion may go in two directions: given an article, tracing its citing articles via citation links is known as forward citation expansion, whereas tracing articles on the reference list of the given article is known as backward citation expansion. The expansion process can be iteratively applied to a growing set of articles (C. Chen, Lin, & Zhu, 2006).

Until recently, the ability to perform citation expansions is essentially limited to the few who have unrestricted access to a citation database that is large enough to sustain iterations of citation expansion as much as needed. For example, the Web of Science used to offer a citation map, which visualized one-step citation expansions forward, backward, or both in a hyperbolic visualization. We explored citation expansions with CiteSeer based on a local copy of CiteSeer, but a cross-disciplinary platform that may cover a wider range of scientific literature was not feasible because of the lack of unrestricted access to a large citation database (C. Chen et al., 2006). We have also learned from our earlier experience with citation expansion that one may benefit the citation indexing approach in several practical and commonly encountered scenarios. For example, we may start with a classic article on a topic of our interest and collect subsequently published articles that have further developed the ideas in the classic article. We may also start with a recently published article and want to find what has been done in the past that has led to the recent publication. We may want to perform citation expansions in both ways, construct a comprehensive collection of publications, and then apply science mapping or visual analytic tools to the most representative dataset and reveal the entire evolutionary history of a subject. We may even study the citation impact of an individual researcher, an institution, a country, or a field of study.

The access to citation data has been improved considerably in recent years. The number of citation data resources is growing. In addition to the long established ones such as the Web of Science, Scopus, Google Scholar, and Microsoft Academic, open access to citation data has increased fast, notably through the effort and the provision of Crossref, Dimensions, and the Initiative for Open Citations[1].

The Initiative for Open Citations (I4OC) promotes the unrestricted access to scholarly citation data in terms of three groups of desirable goals. Specifically, I4OC advocates that citation data should be *structured* so that they are accessible in machine-readable forms for computer programs. I4OC also suggests that citation data should be *separable* from the source product from which citation instances are originated. Finally, and perhaps most importantly, citation data should be *open*, which means freely accessible and reusable.

Dimensions is a major addition to the existing citation data resources. In addition, Dimensions is envisaged to provide a next-generation platform to increase the discoverability of scholarly publications. Dimensions is accessible through an API as well as through its interactive interface with a web browser. The new platform opens new opportunities, including unrestricted access through an iterative process of citation expansion, which we refer to it as cascading citation expansion (CCE). Although performing citation expansion within a large commercial citation database is always a technically viable option for the few with unrestricted access to such databases, such technical capabilities have been essentially beyond the reach of the vast majority of scientists, researchers, students, and practitioners. Therefore, open citation platforms such as Dimensions have the potential to crystalize the resources, services, and enabling tools needed not only for a more efficient and flexible access to scientific literature but also for the development of insightful meta-knowledge so as to better capture the status of scientific knowledge at various levels of granularity.

Searching for accurate and reliable indicators of research performance has a long and often controversial history (L. Leydesdorff, 2008; Waltman, 2016). The increasingly accessible citation data and citation-based performance or excellence indices have exposed biases, pitfalls, distortions, and other types of risks and uncertainties (C. Chen & Song, 2017; Hicks, Wouters, Waltman, Rijcke, & Rafols, 2015). In parallel to the development of indicators of research impact, researchers have increasingly turned to analytic studies of scientific literature to reveal the significance of landmark discoveries and pivotal contributions in the evolution of a scientific field. Science mapping, in particular, has been accepted by a growing number of researchers across numerous disciplines (C. Chen, 2017). The increasing popularity of science mapping studies is due to the instrumental role of science mapping and visual analytic tools (Cobo, López-Herrera, Herrera-Viedma, & Herrera, 2011), notably widely used ones such as VOSviewer (Eck & Waltman, 2010) and our own tool CiteSpace (C. M. Chen, 2004, 2006; C. M. Chen, Ibekwe-SanJuan, & Hou, 2010). Many individual users of science mapping tools typically rely on searching by topic, by journal, or a combination of various aspects of bibliographic metadata. Citation expansions can be performed on commercial and proprietary platforms through functions such as the Citing Articles function of the Citation Report in the Web of Science. For instance, we applied this method in our previous studies of research in regenerative medicine and rare diseases (C. M. Chen, Dubin, & Kim, 2014a, 2014b).

In this article, our goal is to demonstrate how cascading citation expansion tasks can be integrated with the workflow of science mapping and visual analytic studies of scientific literature. First, we will outline a few use case scenarios in which one can explore scientific literature in a wider but still highly relevant context with the help of citation expansion. Then we will illustrate through visualizations generated in CiteSpace what types of information are added as a result of citation expansion and their implications on a visual analytic study of a scientific field. Finally, we will discuss some of the practical issues and remaining challenges. We expect such demonstrated examples will be useful for analysts and curators and distributors of open citation data.

---

[1] https://i4oc.org/

## Cascading Citation Expansion

The idea of citation expansion is a natural extension of citation indexing, which can be considered as an incremental search strategy (E. Garfield, 1963). Given an initial set of publications *S*, we denote the expanded set of publications that directly cite the members of *S* as $Citing^1(S)$. This type of citation expansion is known as forward citation expansion because articles in the expanded set are typically published more recently than members of the *S*. For each expansion, the initial set is called the source set and the resultant set of articles is, of course, called the result set. The result set of one expansion can serve as the source set for the next round of citation expansion, i.e. $Citing^1(Citing^1(S)) = Citing^2(S)$. We will obtain a new set $Citing^n(S)$ if we successively apply the expansion operation for *n* times, which defines a *n*-step cascading forward citation expansion. Expanded sets at different steps of a citation expansion chain may contain overlapping articles because for a given pair of articles <source, target> there may exist alternative citation paths such as source ← target, source ← intermediate ← target, and source ← … ← target. Duplicated articles and overlapping paths can be subsequently removed.

As illustrated in Figure 1, citation expansions can extend the reference set of a single article or a set of articles *S*. Citation expansions of this type are known as backward citation expansion, denoted as $CitedBy^1(S)$ for a given source set *S*. We can successively apply the backward citation expansion for an arbitrary number of steps or until the result set satisfies a predefined requirement, for example, when the expansion has reached publications in the 1940s or when the h-index of the result set exceeds 100.

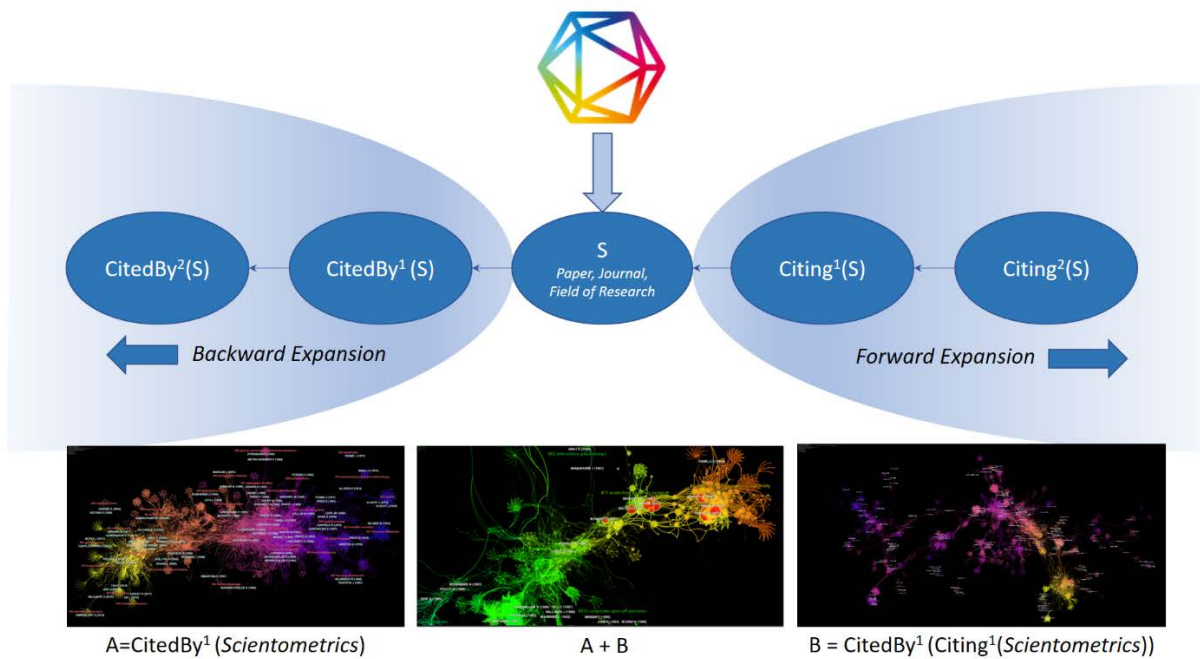

*Figure 1. Cascading citation expansions with Dimensions.*

An immediate benefit of citation expansion is that we can improve the coverage of the set of articles we start with. A broadened coverage is particularly useful when we look for relevant articles that are published in more recent years or in disciplines that we are not familiar with or simply unaware of. We expect that citation expansions will improve the recall of relevant articles considerably while maintaining the stability of the precision to a reasonable extent. As illustrated in Figure 1, one can take an expanded set of articles and visualize the structure and dynamics of a scientific field in a broader context than what the initial source set can offer.

In fact, there are numerous ways to utilize the comparative capabilities enabled by cascading citation expansion. Here are some potential use case scenarios:

1. Given a classic article, find high-impact articles that cited the classic article, directly and indirectly, up to date. For example, we are familiar with the significance of latent semantic indexing through a widely cited 1990 article (Deerwester, Dumais, Furnas, Landauer, & Harshman, 1990) and we would like to explore significant advances since then and preferably trace their connections all the way up to date.
2. Given a journal of our interest, Scientometrics, we are interested in collecting scientometric studies published elsewhere and we would like to study pathways, if any, that connect the core of scientometric research and its applications across a diverse range of disciplines. Furthermore, we would be interested in tracing the potentially foreign origins of concepts, principles, or methods that we have been taken for granted in scientometrics research.
3. Given a field of research, such as Neurosciences, we are interested in where and how it interacts with other disciplines but we would also like to limit the size of our search space whenever possible.
4. Given two research topics or two bodies of research defined by researchers from two institutions, we would like to find citation chains that connect the two bodies of scientific literature or longer alternative pathways that may lead to new hypotheses and discoveries in the spirit of spreading activation (Neely, 1977) and various strategies developed in literature-based discovery (Swanson, 1986).

Generally speaking, in order to show how one set of information differs from the other, one can use an intuitive method in the context of network visualization – a network overlay. A base network serves as the background and the overlay layer serves as the foreground as shown in Figure 2. In particular, overlaying the network derived from the original source set on the network derived from the expanded set is useful to identify areas that are introduced by the citation expansion as well as areas that are core to the original set.

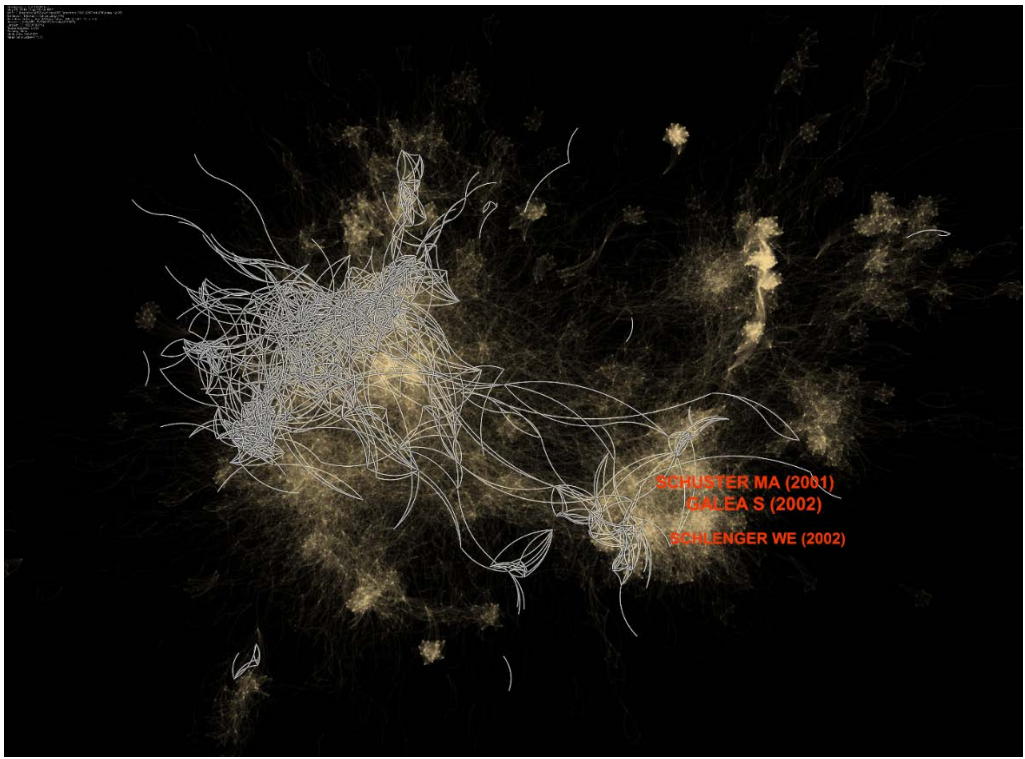

*Figure 2. A network overlay can be used to assist comparisons between a background network and a foreground network.*

## Dimensions

Dimensions provides a web-based platform for users to search for publications, grants, patents, and clinical trials through the Dimensions Search Language (DSL). DSL resembles the commonly used QSL. By default, the user receives the first 20 records. The maximum number of records per page is 1,000 records. It is possible to page through a longer list of search results by setting an offset from the beginning of the list. Dimensions offers an API for programmatically retrieving records of interest.

A publication record in Dimensions typically includes the title of the publication, its authors and their affiliations, the primary field of research (FOR), times cited, the date of publication, the journal title, and a few other bibliographic fields. Dimensions identifies a variety of entities with platform-wide unique id's for entities such as publications, journals, authors, grants, and patents. The publication id, for example, can be used not only to locate the corresponding record but also to identify its citing articles, i.e. articles that cite this particular publication in their references. A recent comparative study between samples from Dimensions and Scopus concluded that the coverage and citation counts of Dimensions are comparable to those of Scopus (Thelwall, 2018).

In addition to the times_cited field in a publication, Dimensions provides other metrics such as altmetric attention score in the altmetric field and the relative citation ratio in the relative_citation_ratio field. Just as a publication may be never cited, a publication may have no altmetric scores. Figure 3 illustrates the publication that has the highest altmetric score in Dimensions. This article in *Scientific American* has been tweeted or retweeted over 14,000 times on Twitter. In contrast, it has been cited only 18 times for a publication with such a high profile on social media. Different metrics are likely to highlight different aspects of publications.

*Figure 3. The publication with the highest altmetric attention score in Dimensions.*

The following examples outline a few potentially common use case scenarios for citation expansion in terms of how one can construct the initial source set of publications. This is probably by far the most common situation a researcher may encounter. We can start our search process with a keyword such as "Latent Semantic Indexing" or "Climate Change". Another common use case starts with a specific publication based on its DOI or its publication id in Dimensions. As shown in Figure 4, a search by DOI returns a record of the 2006 paper on CiteSpace. It has been cited by 554 publications indexed in Dimensions and we can retrieve these citing articles based on the id of this article "pub.1001131784".

```
▼ { 2 items
    ▼ "publications" : [ 1 item
        ▼ 0 : { 4 items
            "title" :
            "CiteSpace II: Detecting and visualizing emerging trends and transient patterns in scientific literature"
            "doi" : "10.1002/asi.20317"
            "id" : "pub.1001131784"
            "times_cited" : 554
          }
      ]
    ▼ "_stats" : { 1 item
        "total_count" : 1
      }
  }
```

*Figure 4. The result of a search by DOI.*

Table 1 shows 17 citing articles to the 2006 CiteSpace paper containing the phrase "emerging trends" in their titles. In one instance, the two words in the phrase are separated. The DSL query for this example is as follows:

```
search publications in title_only for "emerging trends" where references
= "pub.1001131784" return publications[title+year] sort by year
```

*Table 1. 17 citing articles to the 2006 CiteSpace paper containing the phrase "emerging trends" in their titles.*

| Year | Title |
|---|---|
| 2018 | **Emerging trends** and new developments in information science: a document co-citation analysis (2009–2016) |
| 2018 | Mapping hotspots and **emerging trends** of business model innovation under networking in Internet of Things |
| 2018 | The structure and **emerging trends** of construction safety management research: A bibliometric review |
| 2017 | Knowledge Domain and **Emerging Trends** in Organic Photovoltaic Technology: A Scientometric Review Based on CiteSpace Analysis |
| 2017 | An exploration to visualise the **emerging trends** of technology foresight based on an improved technique of co-word analysis and relevant literature data of WOS |
| 2017 | A scientometric review of hotspots and **emerging trends** in additive manufacturing |
| 2017 | Exploring evolution and **emerging trends** in business model study: a co-citation analysis |
| 2017 | **Emerging trends** and new developments in monoclonal antibodies: A scientometric analysis (1980–2016) |
| 2016 | **Emerging Trends** and New Developments on Urban Resilience: A Bibliometric Perspective |
| 2016 | A review of **emerging trends** in global PPP research: analysis and visualization |
| 2015 | A scientometric review of **emerging trends** and new developments in recommendation systems |
| 2014 | **Emerging trends** and new developments in regenerative medicine: a scientometric update (2000 – 2014) |
| 2013 | Collective dynamics in knowledge networks: **Emerging trends** analysis |
| 2012 | **Emerging trends** in regenerative medicine: a scientometric analysis in CiteSpace |
| 2010 | **Trends** in research foci in life science fields over the last 30 years monitored by **emerging** topics |
| 2009 | Bibliometrics for advancement R&D Planning : Detecting **Emerging Trends** in Scientific Literatures |
| 2008 | Detecting and Visualizing **Emerging Trends** and Transient Patterns in Fuel Cell Scientific Literature |

The initial source set of articles can be easily defined by a journal of interest, for example, *Scientometrics* or *Nature*. If we know the ISSN number of a journal, it can be used in a query as in the following example, which specifies that we would like to find articles published in the journal Scienometrics with at least one citation and its own reference list is not empty:

```
search   publications   where   issn   in   ["0138-9130",   "1588-2861"]   and
times_cited>0 and references is not empty return publications[all]
```

Another point of entry uses the field of research (FOR). FORs are initially trained by machine learning algorithms and verified by human experts (Herzog, Sorensen, & Taylor, 2017). For example, the following query searches for all publications in neurosciences and the results are sorted by altmetric attention scores. A total of 1,614,810 publications on Dimensions are qualified. The publication with the highest altmetric score of 3,023 has been cited 359 times.

> search publications where FOR.name="1109 Neurosciences" return publications[title+times_cited+altmetric+year] sort by altmetric limit 1

Table 2 illustrates the number of top-sliced publications at several levels of citations in different disciplines with different aggregation units. At the highest scale, among the 94.6 million publications in the entire Dimensions, there are three articles with citations over 100,000 times each and an h-index of 3,013. At the next level, the field of neurosciences has 1.6 million publications and an h-index of 1,043. *Nature*, as a single journal alone, has 641,063 publications and an h-index of 1,403, which is higher than the field of neurosciences. The *Scientometrics* journal has 5,093 publications and an h-index of 89. Finally, at the level of individual publications, the 1990 classic paper on Latent Semantic Indexing has been cited 4,744 times and an h-index of 125, whereas the 2006 landmark paper for CiteSpace has 554 citations and an h-index of 36.

These numbers are useful for us to estimate the volume of publications involved at a particular step of a citation expansion. As we can see, the number of publications increases sharply as we move from a higher threshold to a lower one. For example, there are 2,746 articles at the 1,000-citation level. In comparison, there are 57,655 qualifying publications at the 100-citation level, approximately increased by 20 times. If we use citation levels to control the scope and scale of a citation expansion, citation expansions at the 100-citation level will need to process 20 times more records than at the 1,000-citation level. Such configurations may directly influence the time taken to complete the expansion process.

*Table 2. The volume of publications at various levels of citations (Source: Dimensions, as of May 28, 2018).*

| Seed Article Set | Size | Cites ≥1 | ≥10 | ≥$10^2$ | ≥$10^3$ | ≥$10^4$ | ≥$10^5$ | h-index |
|---|---|---|---|---|---|---|---|---|
| All publications in Dimensions | 94,578,660 | 47,809,718 | 19,752,566 | 1,584,237 | 25,452 | 266 | 3 | 3,013 |
| Field of Research: 1109 Neurosciences | 1,614,809 | 1,357,321 | 826,244 | 97,844 | 1,173 | 2 | 0 | 1,043 |
| *Nature* jour.1018957 | 641,063 | 312,538 | 170,138 | 57,655 | 2,746 | 15 | 1 | 1,403 |
| *Scientometrics* jour.1089056 | 5,093 | 4240 | 1803 | 72 | 0 | 0 | 0 | 89 |
| Citing(LSI) pub.1012153938 | 4,744 | 3600 | 1298 | 160 | 10 | 0 | 0 | 125 |
| Citing(CiteSpace) pub.1001131784 | 554 | 365 | 142 | 11 | 0 | 0 | 0 | 36 |

## CiteSpace

CiteSpace is a freely available Java application for visualization and analytic studies of scientific literature (C. Chen, 2017; C. M. Chen, 2004, 2006; C. M. Chen et al., 2010). Since it was made freely available in 2004, it has undergone major revisions and frequent updates to support an increasing number of sources of citation data, notably including the Web of Science, Scopus, Crossref, and most recently Dimensions.

A typical application of CiteSpace consists of the following steps:
- Data Collection
- Visualization: Geographic Overlay (Optional)
- Visualization: Dual-Map Overlays (Optional) (C. M. Chen & Leydesdorff, 2014)
- Network Modeling
- Interactive Network Visualization: Cluster View, Timeline View, Timezone View
- Structural Variation Analysis (Advanced) (C. M. Chen, 2012)

The major visual analytic path in CiteSpace starts with the construction of a set of publications. CiteSpace supports document co-citation analysis, collaborative network of coauthors, network analysis of keywords and noun phrases extracted from the titles and abstracts of the source articles. Examples included in this article illustrate the co-citation networks of cited references. Given a set of source articles, CiteSpace slices the time interval of interest into a series of time slices such that articles published in each time slice form a network. Articles in the entire source set thus generate a series of networks. These networks are synthesized further to form a panoramic view of the subject domain.

The synthesized network is then divided into clusters based on the strength of co-citation links. The nature of each cluster is summarized by a list of representative terms, which can be extracted from titles, abstracts, or author provided keywords of the citing articles of the corresponding cluster. These representative terms are used to characterize the nature of each cluster from the perspective of its citers. In other words, cluster labels reflect the influence of a cluster in the context of how they are cited in subsequently published articles, which may differ from the original publications. For more details, please refer to our previous studies (C. Chen, 2017; C. M. Chen et al., 2010).

A more advanced and perhaps less commonly used analytic approach supported by CiteSpace is structural variation analysis (SVA). SVA was originally proposed in (C. M. Chen, 2012) in order to provide a theoretical and experimental framework for how we may identify new publications in terms of their transformative potential – the potential that they may profoundly alter the macroscopic structure of a field of study. In essence, the idea is based on the assumption that transformative contributions can be detected by the changes introduced by a particular publication in the underlying network. This approach does not rely on the number of citations received by the particular publication. Instead, the approach is based on the degree to which how newly added citation links may change the structure between clusters and how much differ from the network prior to the publication of the article. The quality of SVA is affected by the quality of the underlying network that serves as a reference point for each year of the window of the analysis. Thus, we expect citation expansions will improve the quality of SVA because citation expansions enrich the baseline network and extend the coverage.

## Illustrative Use Cases

We illustrate some common use cases using forward and backward citation expansion.

*Forward Citation Expansion: Latent Semantic Indexing (1990-2017)*

The forward expansion example illustrates a common scenario in research. It is quite common that all we have to start with is a well-written, clearly relevant, and highly-cited classic paper. Such classic papers tend to be published a long time ago. How can we find out what happened ever since? With Eugene Garfield's idea of citation indexing, tracing subsequent publications along the direction of forward citation expansion is conceptually well-known but practically still not quite straightforward to carry out.

We take the 1990 paper on Latent Semantic Indexing (Deerwester et al., 1990) as the starting point of a series of forward citation expansion. Each time as we take one citation expansion forward, more recently published articles will be retrieved from Dimensions and added to the growing collection of articles. These articles are relevant to the original topic of Latent Semantic Indexing through an increasingly longer chain of citation. In this way, the timespan of the articles extends from the initial year of 1990, which is the year of the seed paper, to 2017. To illustrate the expansion process, we did not configure for the most comprehensive possible capture. In this example, we limit the rate of expansion to 10 per article. For each article in the current generation, up to 10 of its citing articles that are most cited themselves will be retained for the next expansion. In other words, the increase of the population with each step of expansion is limited to 10 times of the current generation's population – with a k-step forward expansion the population is controlled under $10^k$. This rate limit can be configured by the user empirically to suit the analytic need. More sophisticated selections are possible, for example, based on the h-index of the next generation, citations normalized by corresponding fields of research and years since publication, and other metrics readily available in Dimensions, such as altmetric attention scores and relative citation ratios.

Figure 5 shows the largest connected component of the network resulted from the forward citation expansion. The final range of the additional articles is between 1990 and 2017. The largest component contains 2,502 references cited by the expanded set, which represents 70% of the entire network of references cited by the highly selective expansion process. The modularity of the network is 0.8515. Different topic areas are noticeably separated from one another, which makes it easier for visual exploration.

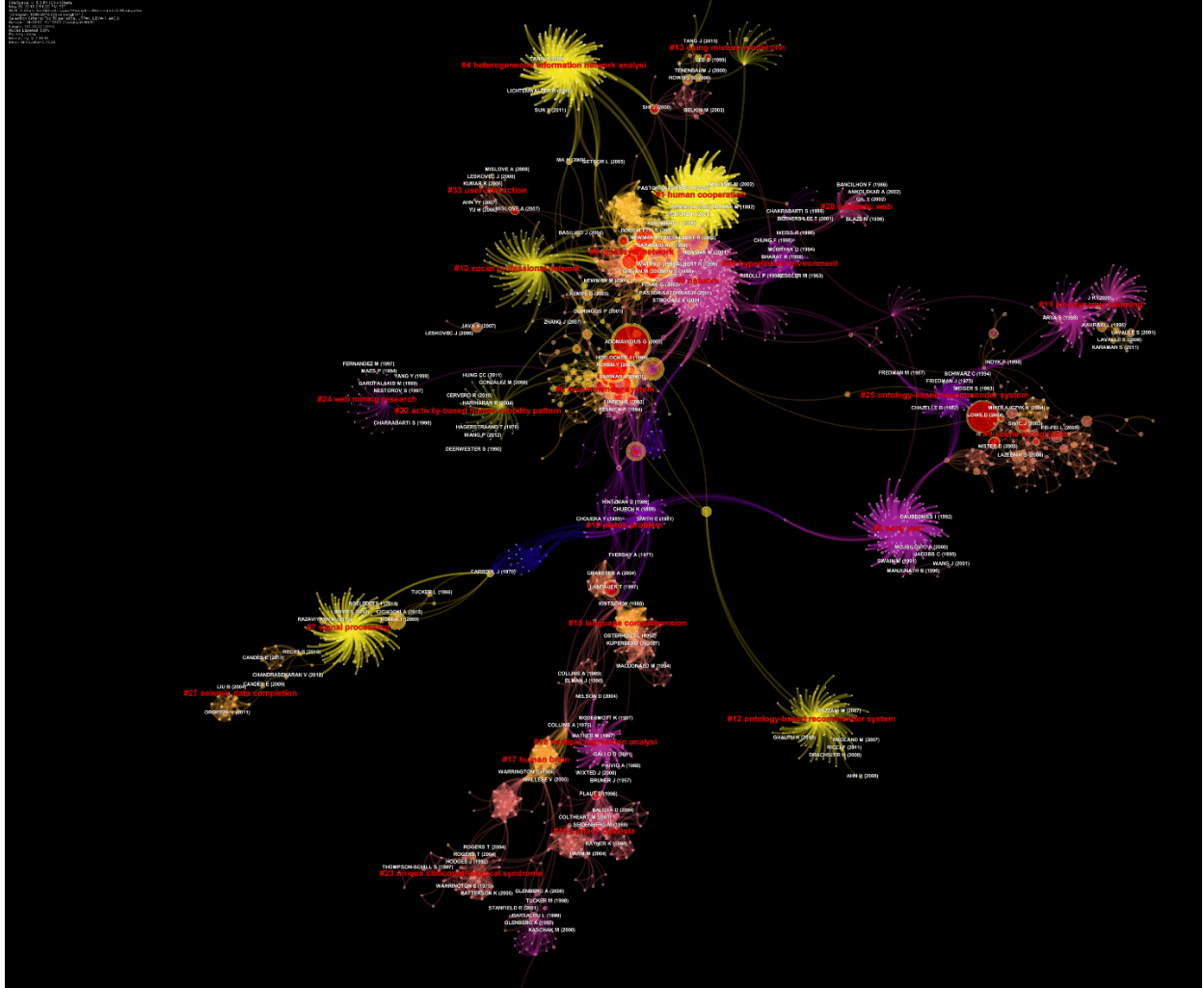

*Figure 5. An overview of the forward citation expansion from the 1990 article on Latent Semantic Indexing (1990-2017).*

The 1990 seed article is marked by a star in the center of the network resulted from a series of forward citation expansions from 1990 until 2017. In Figure 6, the cluster below the seed article is labeled as #19 Plato's problem, which is the predeceasing generation of the seed article, whereas the cluster above the seed article, #6 recommender system, is one of the next generations of articles that are relevant to the topic of LSI through citation chains. The key papers in this cluster appeared about 10 years after the LSI seed paper and many of them have shown citation bursts, i.e. citation tree rings in red. For example, (Koren, Bell, & Volinsky, 2009) on matrix factorization has a strong citation burst of 17.66. (Linden, Smith, & York, 2003) on Amazon.com recommendations has a citation burst of 8.13. (Herlocker, Konstan, Borchers, & Riedl, 1999) on collaborative filtering has a citation burst of 5.02. (Sarwar, Karypis, Konstan, & Riedl, 2001) on item-based collaborative filtering and (Resnick, Iacovou, Suchak, Bergstrom, & Riedl, 1994) on GroupLens are highly cited members but no citation bursts detected.

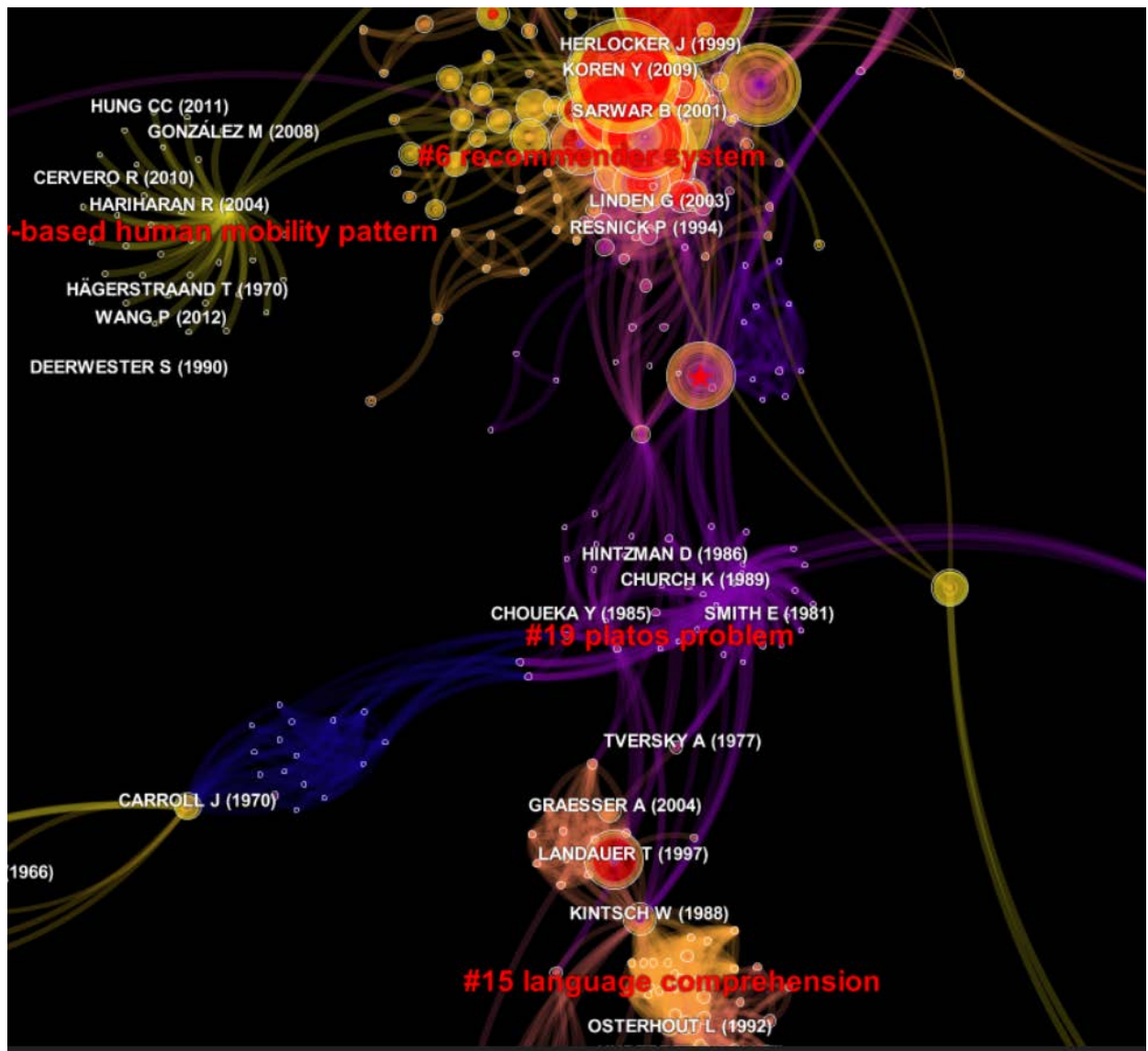

*Figure 6. Neighboring clusters of the original seed article on LSI (the node marked by a red star).*

If we move upwards from the recommender system cluster, we will encounter the largest cluster #0 network and #5 multilayer network (Figure 7). Based on the key members in the two clusters, they represent the study of complex networks, small-world networks (Watts & Strogatz, 1998), and scale-free networks (Barabási, Albert, & Jeong, 1999). The cluster #5 multilayer network contains relatively more recent publications, including community structure detection (Girvan & Newman, 2002) and the structure and function of complex networks (Newman, 2003). Clusters on the right are related to hypertext and the semantic web.

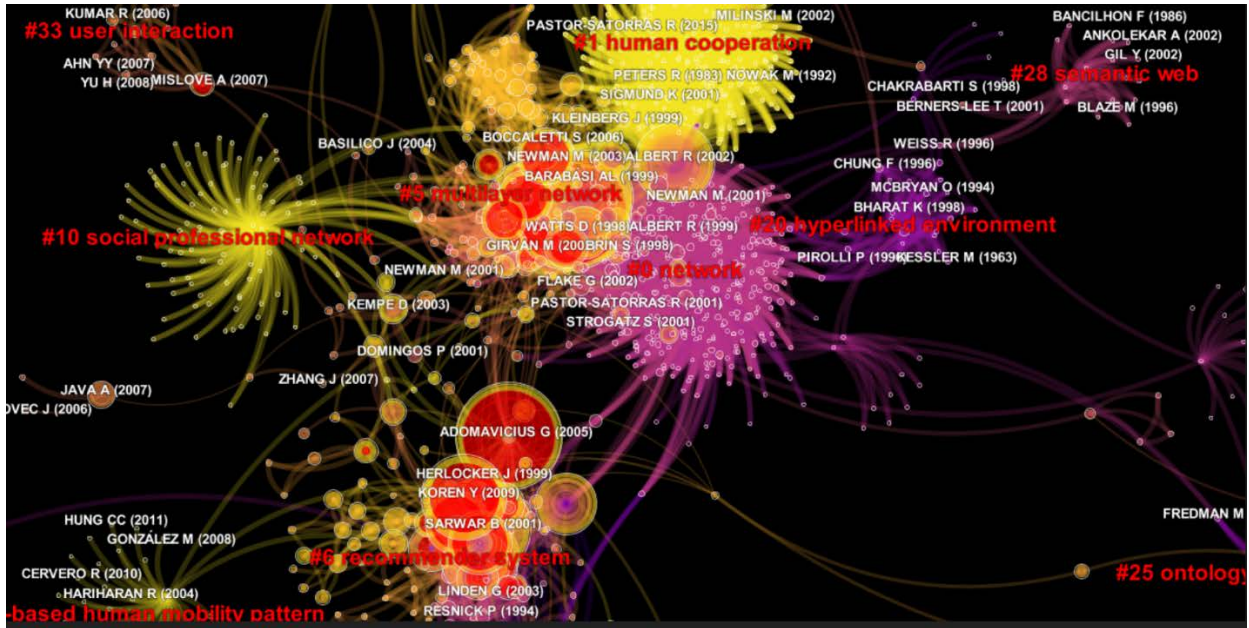
*Figure 7. Moving upwards from the recommender system cluster.*

In CiteSpace, the user can step through the visualized network over time one year at a time to inspect the concentration of research activities in a particular year. For example, Figure 8 shows a snapshot of the status of the network in 2017. As we can see from time-stamped frames of such snapshots, by 2017, active areas have drifted away from the 1990 LSI article on LSI.

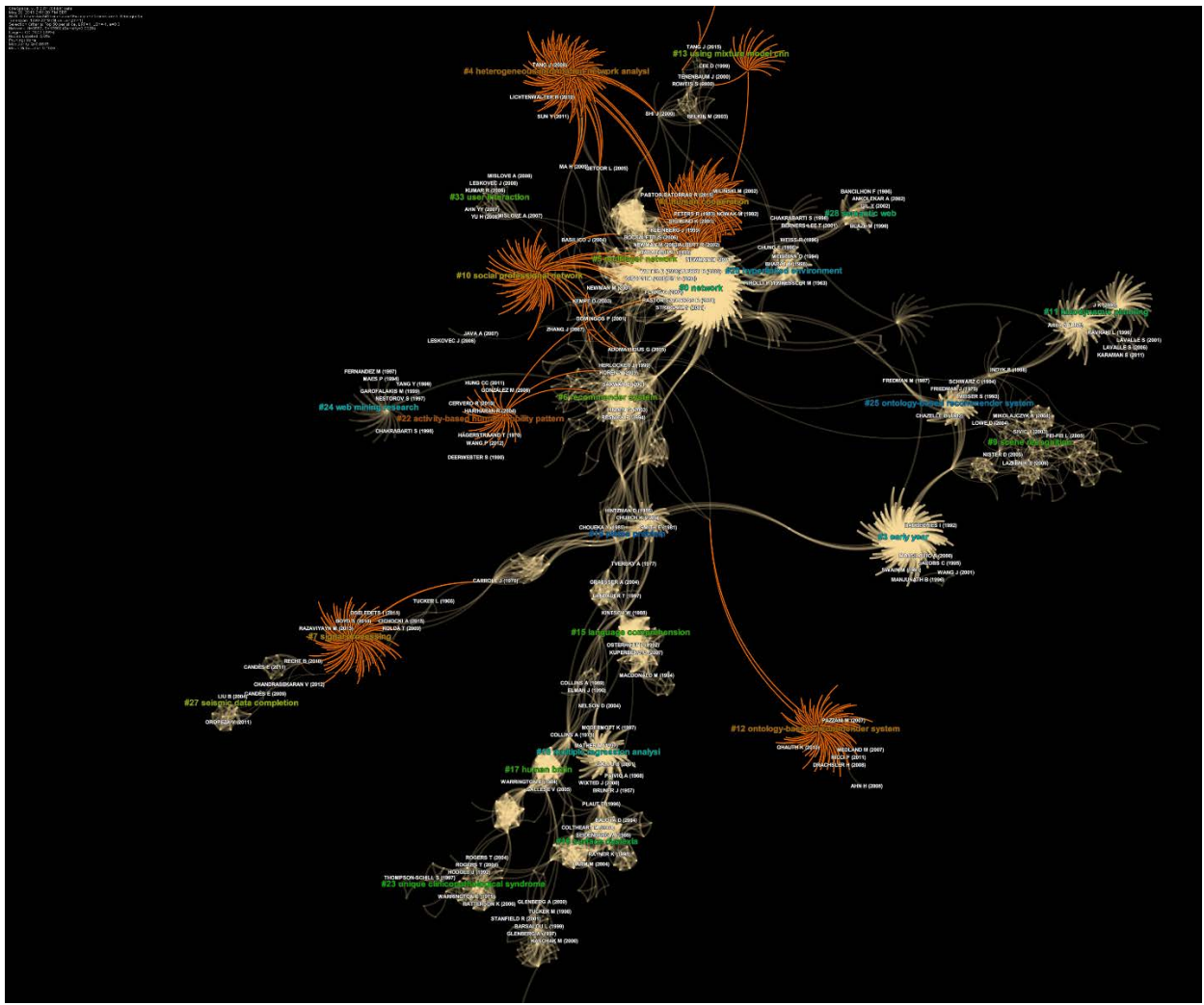

*Figure 8. A snapshot of co-citation links added in 2017.*

The example of the forward citation expansion from the 1990 LSI article illustrates how we can catch up the research that takes place since the publication of a seed paper or even a set of seed papers. More importantly, since these newly found articles are relevant to the original topic, we can examine the conceptual evolution with a desirable degree of continuity. Such continuity may not be necessarily preserved if we simply retrieve a bulk of publication records from a citation data source as vast as Dimensions. Thus, we emphasize the importance to preserve the connectivity through citation expansion and recommend this strategy in similar situations.

*Backward Citation Expansion: Scientometrics – A Galactic View*
To illustrate the usefulness of backward citation expansion, we use the journal *Scientometrics*, i.e. all its publications found in Dimensions, as the initial set. Applying backward citation expansions successively to the result set from the previous round of expansion will expand the citation space by adding more and typically articles published in earlier years.

Figure 9 shows a co-citation network based on publications in *Scientometrics*. The colors based on a perceptually uniformed color map depict the time of publication. The earliest links are in the right-most area of the display, whereas the most recent ones are in the left-most area of the display. The quality of the references is at least as good as its counterpart and long-established Web of Science. Given its open access accessibility, this is a very encouraging and promising route to pursue, especially for many who may not have an unrestricted access to science citation data.

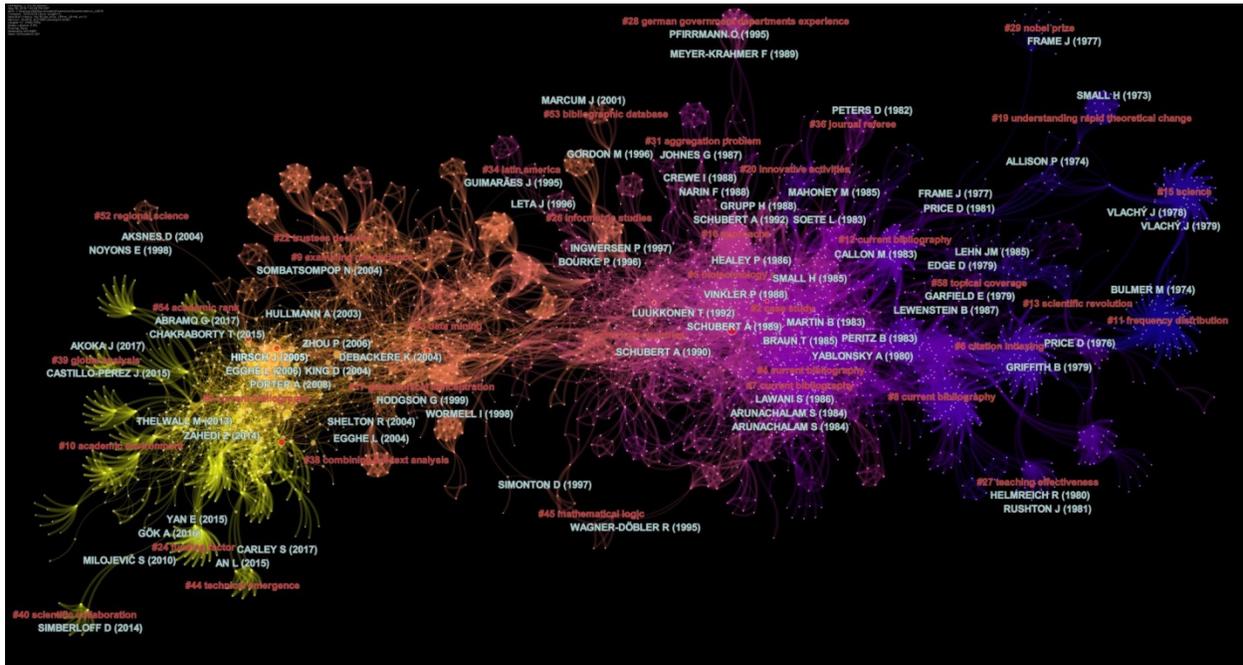

*Figure 9. A galactive view of references cited by articles published in the journal Scientometrics. Data Source: Dimensions.*

The backward citation expansion takes the Scientometrics publications and trace each reference cited in the dataset. The cited reference is then used to drill down further to find its own references. The process may repeat until a set of predefined conditions for termination are met. In this example, the original timespan is between 1978 and 2018. The time interval of the expanded set is extended to 1950.

The expanded set is further normalized by fields of research and the number of years since publication. Articles with citations above the median of the citations in their own fields, i.e. the $50^{th}$ percentile of the field-normalized citations, are used to as the input of the visual analytic study. Articles with citations below the field-specific citation median are not used for further study. Field normalization is particularly useful when a diverse range of disciplines are involved so that each discipline can be appropriately represented along with disciplines that may manifest their citations at significantly different scales (Loet Leydesdorff, Bornmann, Mutz, & Opthof, 2011).

The backward citation expansion and the field normalization led to a co-citation network of 9,295 references cited by publications between 1950 and 2018. The overview of the network in Figure 10 shows that the earliest research topics are located near the upper left and the lower left regions of the visualization. Areas in the middle of the upper and lower regions correspond to activities in the middle stages of the time interval. Areas starting from the center of the lower center of the network and extending towards the right represent more recent topics, with a shape like a peninsula. We will inspect the following three areas in more detail, namely, 1) the earliest topic areas in the upper left region, 2) the lower middle area, and 3) the peninsula-shaped area.

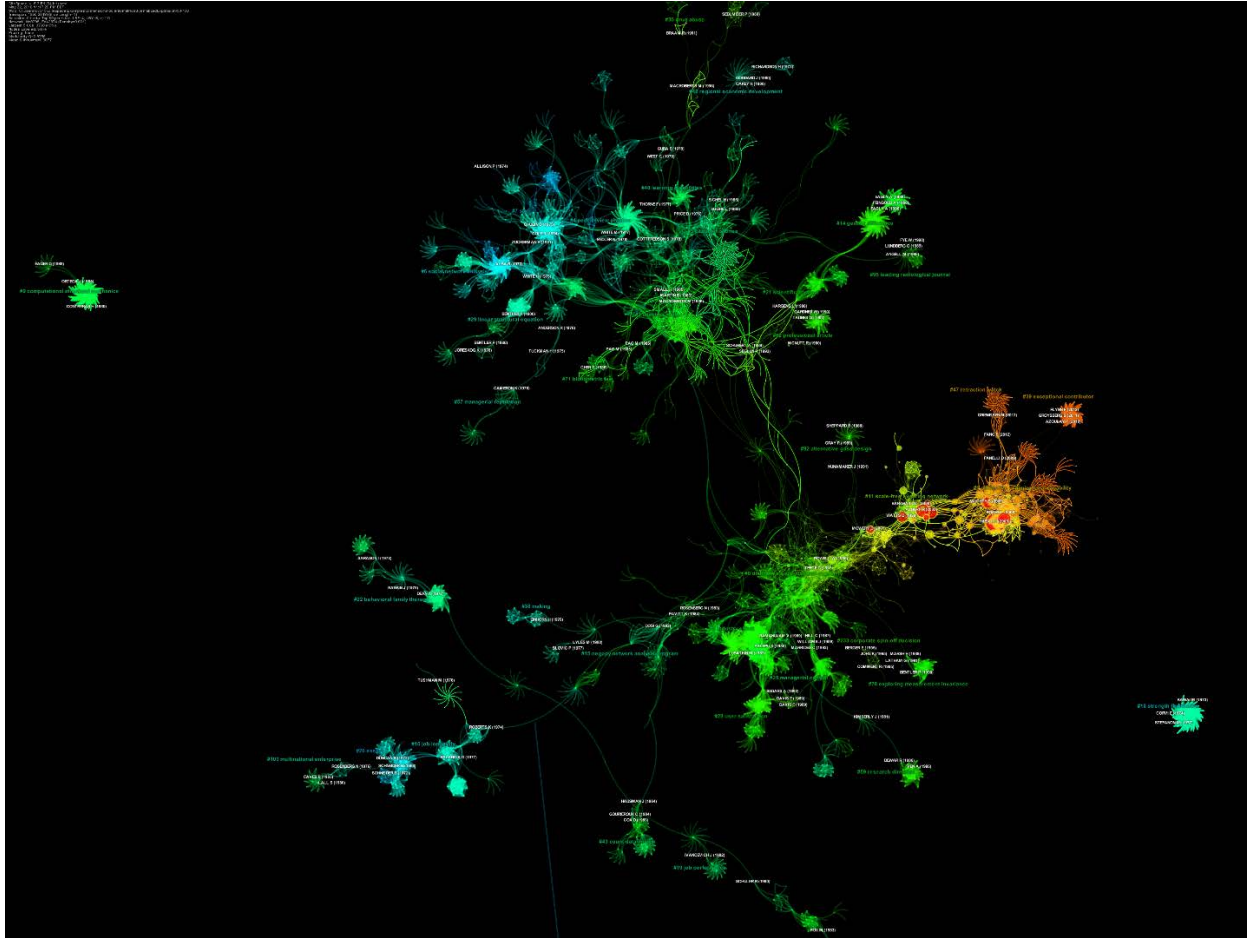

*Figure 10. The network resulted from a field-normalized backward citation expansion of the Scientometrics collection.*

## Inspecting Three Areas in More Detail

The upper left area (Figure 11) contains clusters of some of the earliest references in the window of our study. For example, cluster #6 social network analysis and cluster #4 peer-review practice feature articles published in early 1970s, including (Zuckerman & Merton, 1971) on patterns of evaluation in science and (Cole, Cole, & Beaver, 1974) on social stratification in science. Clusters in the lower right corner of this region include cluster #2 citation matter and cluster #3 citation histories, including (Small, 1985) on clustering with co-citations, (Martin & Irvine, 1983) on assessing basic research, and (MacRoberts & MacRoberts, 1986) on quantitative measures of communication in science. Solid and sharp lines depict co-citation links made articles from the original set. In contrast, soft and blurry lines depict co-citation links made exclusively by the expanded set. In other words, we would have completely missed the clusters #4, #6, #7, #40, and #49 in this region alone and partially missed #2 and #3.

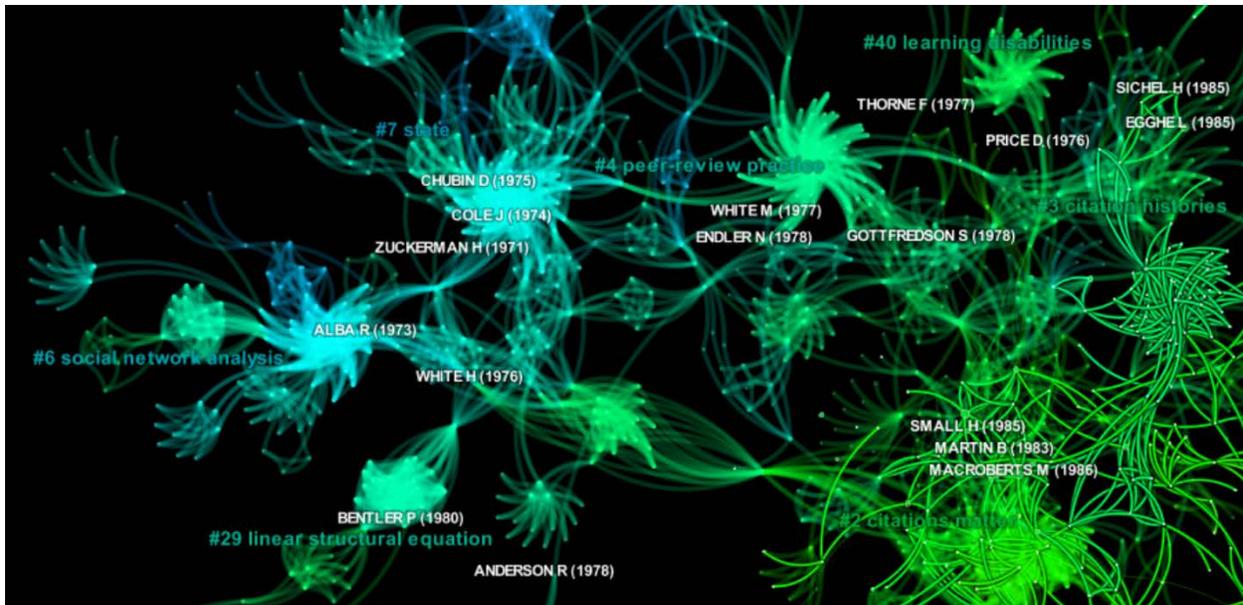

*Figure 11. Branches extending towards the upper left region. The different line styles highlight the effect of the backward citation expansion: strong and sharp lines are from citations made by the original set of publications, whereas soft and blurry lines are resulted from the backward citation expansion.*

The lower middle region of the visualization overview (Figure 12) includes clusters associated with user satisfaction (#28), managerial control (#25), industry fleet (#5), corporate spin-off decision (#233), and research direction (#59). These topics are weakly related to the core of scientometrics. The absence of the solid and sharp lines in this area indicates that this area would have been missed without backward citation expansion. For example, in cluster #5, (Lubatkin, 1987) is on merger strategies and stockholder value and published in Strategic Management Journal. Similarly, (Palepu, 1985) is also from the Strategic Management Journal on diversification strategy and profit performance measures. In cluster #59, (Dewar & Dutton, 1986), published in Management Science, is on the adoption of radical and incremental innovations. Similarly, (Ven, 1986), also published in Management Science, focuses on the management of innovations. These journals are not typically considered as part of the core literature of scientometrics.

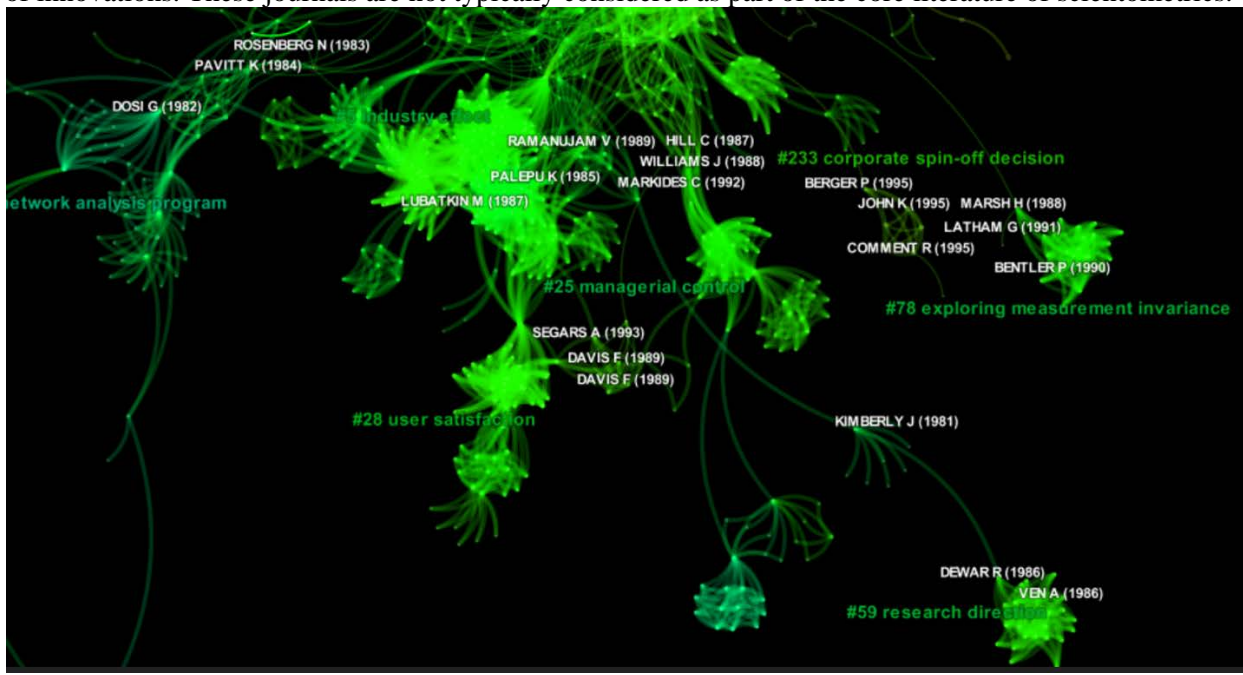

The most active areas are located along the major branches connecting the clusters on the right to the lower middle areas (Figure 13). The largest clusters are also found in this area. For example, the largest cluster #0 distinctive organizational capability is located in the lower left of the closer view of the region. The most cited members of this cluster include (Powell, Koput, & Smith-Doerr, 1996) on interorganizational collaboration and the locus of innovation, (Mowery, Nelson, Sampat, & Ziedonis, 2001) on the growth of patenting and licensing by U.S. universities, and (Teece, Pisano, & Shuen, 1997) on dynamic capabilities and strategic management. The latter two are also found to have strong citation bursts. Publications in this cluster are overwhelmingly from journals such as Administrative Science Quarterly, Research Policy, and Strategic Management Journal.

The structure of cluster #0 reveals that (Teece et al., 1997) and (Powell et al., 1996) are relatively center to the cluster. In contrast, (Mowery et al., 2001) is more like a gate keeper or a boundary-spanning agent because it appears in the middle of the pathways that link cluster #0 on the left and the rest of the peninsula on the right.

Mowery (2001) also divides the references to its left and right. References to its left are essentially connected by soft and semi-transparent lines, indicating they are introduced by the backward citation expansion. In contrast, references to its right are mostly connected by solid and sharp lines, which means they are included in the original network prior to the citation expansion. The first cluster to the right of (Mowery et al., 2001) is cluster #11 scale-free evolving network, featuring high-impact classic papers on small-world networks (Watts & Strogatz, 1998), scale-free networks (Barabási et al., 1999), and a widely cited review of complex networks (Albert & Barabási, 2002).

Moving further towards the east in Figure 13, cluster #1 distinctive organizational capability shares the same label with cluster #0, which is essentially introduced by the backward citation expansion. The fact that the two clusters share a representative label means that the two clusters are formed by some common research topics so that authors cite references in both clusters. Cluster #1 includes (Hirsch, 2005), which introduces the high-impact and often controversial h-index, (Egghe, 2006), which introduces the g-index as an improvement of the h-index, (Wuchty, Jones, & Uzzi, 2007) on the increasing dominance of teams in science, and (Radicchi, Fortunato, & Castellano, 2008) on universal properties of citation distributions. Thus, this cluster clearly focuses on the development of metrics for assessing scientific impact. As the network overlay of the articles from Scientometrics shows, this is the core area of scientometric research.

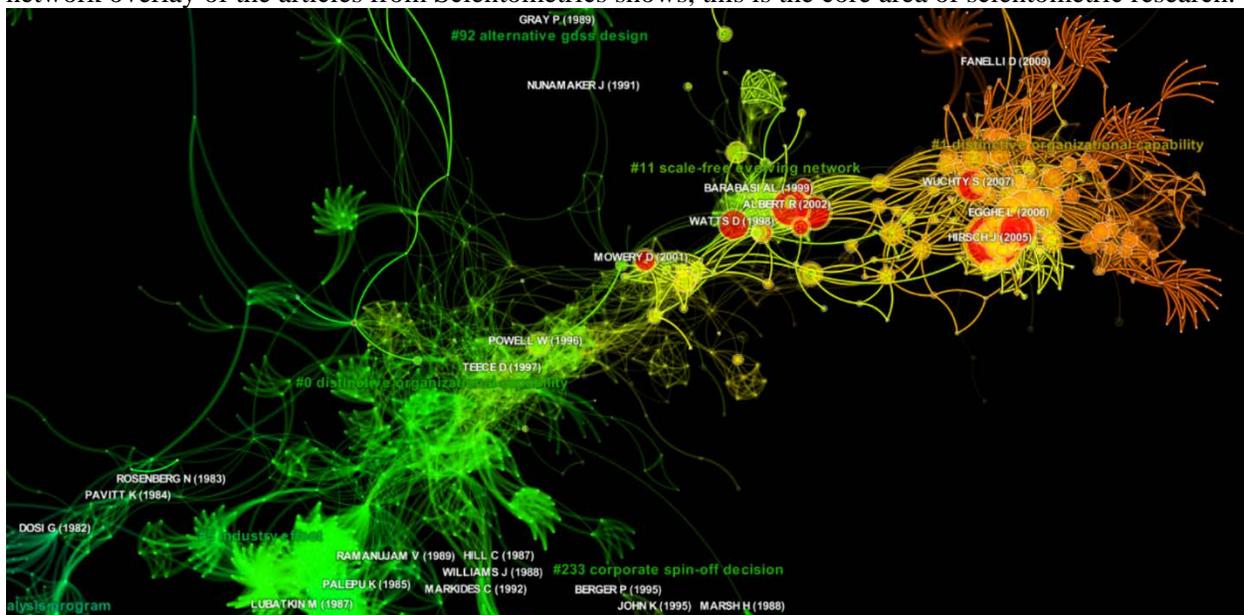

*Figure 13. The mid-right region of the overview display, featuring core areas of scientometrics and expanded areas of innovations and strategic management.*

*Structural Variation Analysis*

We expected that since cascading citation expansions improve the coverage of a citation data set, the improved coverage should in turn improve the quality of Structural Variation Analysis (SVA) because the baseline network is enriched by more relevant references. We compared the results of SVA before and after applying backward citation expansions to the Scientometrics set between 2010 and 2016. Figure 14 depicts the co-citation links made by the top 10 articles that are responsible for the largest modularity change. The dashed lines in red are novel connections made by these articles. The solid lines in purple are connections exist prior to the publication of these articles.

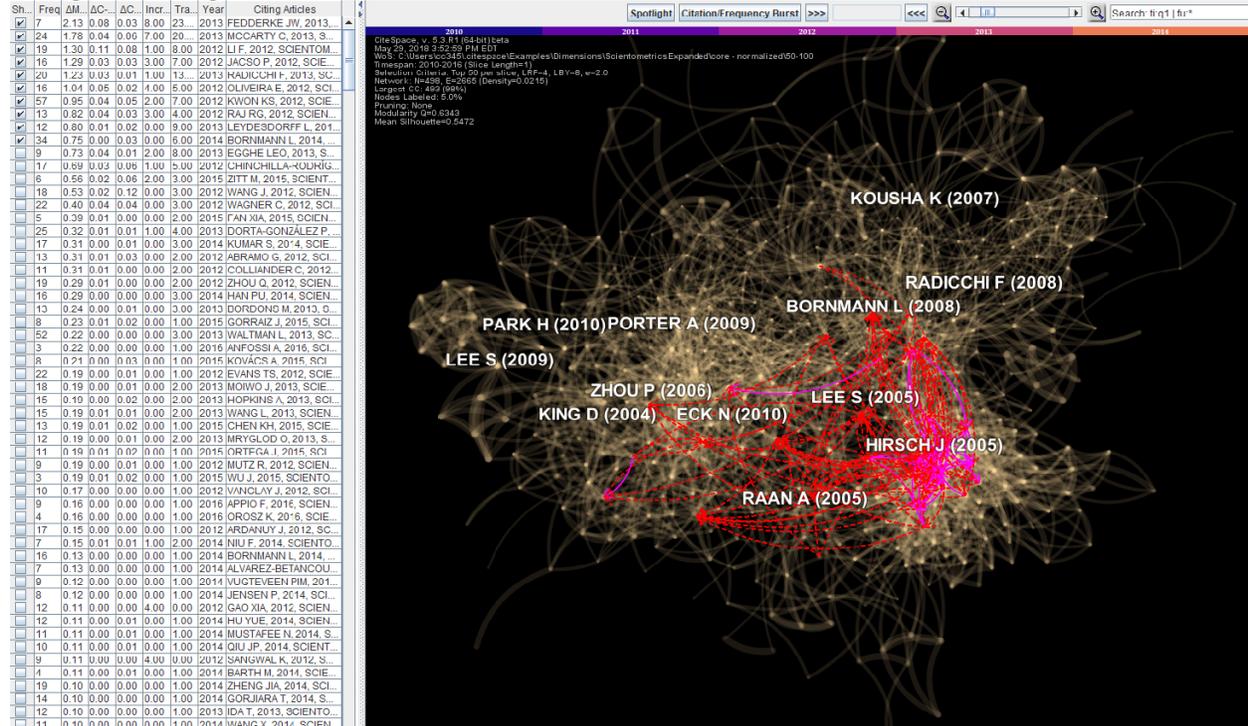

*Figure 14. SVA: Before backward citation expansion. There are 17 clusters. Top 10 articles making the largest modularity changes are shown in dashed lines in red and solid lines in purple.*

Figure 15 depicts the result of SVA after the backward citation expansion. Two articles with stars are not only introducing modularity changing links to the network but also established as cited references that are prominent enough to be part of the network. The two articles are (Radicchi & Castellano, 2012) and (Waltman et al., 2012) on Leiden ranking 2011/2012. In contrast, no article in the pre-expansion set fall into this category.

The article with the highest transformative potential score is (Bornmann & Haunschild, 2016) on how to normalize Twitter counts, whereas no article has scored on this measure. The correlation between the centrality divergence and the global citation count is 0.42 for references with centrality changes, whereas the correlation prior to the expansion was 0.20.

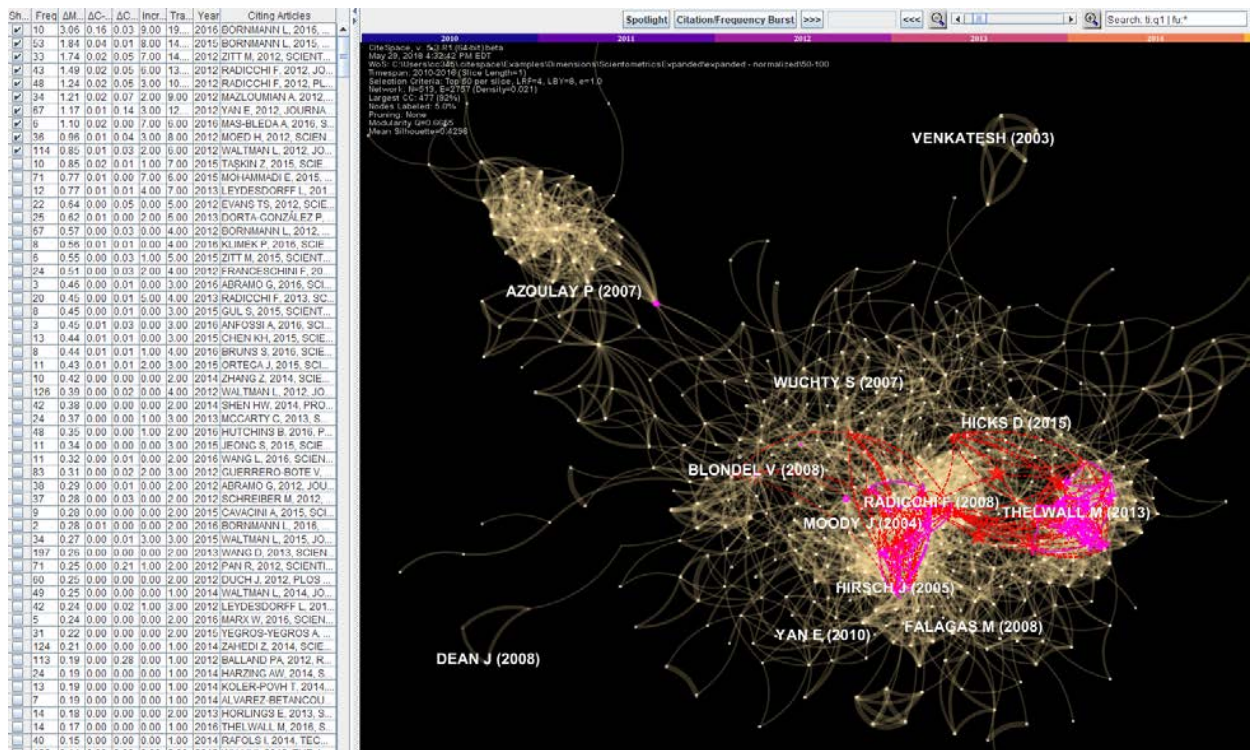

*Figure 15. SVA: After the backward citation expansion, containing 35 clusters. Links made by the top 10 articles responsible for the largest modularity changes are highlighted.*

As shown in the Scienometrics example, backward citation expansions can significantly enrich the coverage of relevant references. This is especially useful if we need to trace to the origin of a concept. Developing a better understanding of where the boundaries are between publications in *Scienometrics* and other relevant but beyond the initial collection of articles is beneficial for researchers to understand the scope of a journal or the scope of a discipline. Our illustrative examples also show promising signs that backward citation expansions may improve the quality of existing methods that rely on the quality of a network of references as a baseline. Our results suggest that Structural Variation Analysis can benefit from such citation expansions with increased correlations between structural variation metrics and the global citations.

## Challenges and Opportunities

Our exploration with Dimensions and newly added functions in CiteSpace has revealed some practical issues that may require the attention of Dimensions and open-access service providers in general. There are hundreds of publications in the journal Scientometrics missing between 2000 and 2003 on Dimensions. In comparison, these missing publications are available in the Web of Science (Figure 16). The reason of the missing data in this particular case is currently unknown. We are sharing the details with the Dimensions team to identify the causes.

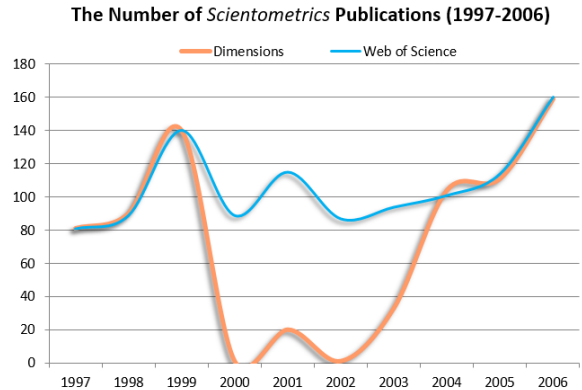
*Figure 16. Missing data from Dimensions.*

We also encountered another type of problems when requesting data from Dimensions. When we retrieved Scientometrics publications from Dimensions, CiteSpace found 79,781 valid and resolvable references. However, we also found 122 references that are incomplete or otherwise invalid. For example, the article with the doi of 10.2307/146070 does not have the year of publication on Dimensions. Although we can resolve it by referring to other open access resources such as Crossref to obtain the missing information, such problems should be at least monitored periodically. In addition, many of the incomplete publications share their prefix of their DOIs, e.g. 10.2307, suggesting this may be a systematic error that should be cost-effective to fix. Currently, publication records in Dimensions do not include abstracts and author defined keywords. Grant records currently do not include information on principal investigators.

We have made extensions in CiteSpace so that the user can visualize scientific literature retrieved through an integrated user interface using DSL queries. In addition, the user can selects altmetric attention scores or relative citation ratio scores to render the visualization of an underlying network of references. Since attention scores measured in terms of altermetrics and traditional citations in scholarly publications may differ considerably, it is useful to be able to explore the scientific literature from a diverse range of perspectives. We are confident that the overall quality of the data on the Dimensions platform will be increasingly improved as more researchers make use of the valuable resources.

## Conclusions

Based on our exploration of implementing forward and backward citation expansions with the Dimensions API and conducting visual analytic studies with new functions added to CiteSpace, we conclude that the Dimensions open access platform for scholarly publications provides a favorable resource for the study of scholarly communication and the development of a scientific field. We also recommend that cascading citation expansions should be adopted whenever possible for the distinct benefits such as bridging the gap between a representative set of sample articles and the state of the art at the present and tracing the intellectual origin of a research topic of interest. Finally, when combining open access resources of citation data with analytic tools such as CiteSpace, researchers and analysts are better informed about the significance of scientific activities. We are able to access not only indicators of the creation and acceptance of scientific contributions but also the developments that set the nature of particular contributions in a broader and meaningful context.

## Acknowledgements

The author acknowledges the support of the National Science Foundation (award number #1633286). Special thank you to Mike Taylor at Digital Science for his persistent and valuable assistance.